\documentstyle[12pt,graphicx,subfigure]{article}
\def\slash#1{\setbox0=\hbox{$#1$}#1\hskip-\wd0\hbox to\wd0{\hss\sl/\/\hss}}
\begin{document}
\baselineskip=20 pt
\def\l{\lambda}
\def\L{\Lambda}
\def\b{\beta}
\def\a{\alpha}
\def\d{\delta}
\def\g{\gamma}
\def\mphi{m_{\phi}}
\def\dnul{\partial_{\nu}}
\def\dnuu{\partial^{\nu}}
\def\dmul{\partial_{\mu}}
\def\dmuu{\partial^{\mu}}
\def\eps{\epsilon}
\def\hphi{\hat{\phi}}
\def\vphi{\langle \phi \rangle}
\def\mph{m_\phi}
\def\etamunu{\eta^{\mu\nu}}
\def\bfl{\begin{flushleft}}
\def\efl{\end{flushleft}}
\def\bea{\begin{eqnarray}}
\def\eea{\end{eqnarray}}
\def\bi{\begin{itemize}}
\def\ei{\end{itemize}}
\def\goes{\rightarrow}

\begin{center}
{\large\bf { Neutral $Z$ boson pair production due to radion 
resonance in the Randall-Sundrum model: prospects at the CERN LHC.\\}}  
\end{center}

\begin{center}
{\large\sl {Prasanta Kumar Das}~\footnote{Present address: Institute of Mathematical Sciences, C.I.T Campus, Taramani,
Chennai~600113, India. Electronic address:
dasp@imsc.res.in}
}
\vskip  5pT
{\rm
Chung-Yuan Christian University, \\
Chung-Li, Taiwan 320, Republic Of China}\\

\end{center}

\vskip 10pT
\centerline{\bf Abstract}

\vskip 5pT
\noindent 
{\it  ~The Neutral $Z$ boson pair production due to radion 
resonance at the Large Hadron 
Collider (LHC) is an interesting process to explore the notion of warped geometry (Randall-Sundrum model). Because of 
the enhanced coupling of radion with a pair of gluons due to trace enomaly and top(quark) loop,
the radion can provide larger event rate possibility as compared 
to any New Physics effect. Using the proper radion-top-antitop 
(with the quarks being off-shell) coupling, we obtain the correct radion 
production rate at LHC and explore several features of a heavier radion 
decaying into a pair of real $Z$ bosons which subsequently decays into charged 
$4~l (l=e,~\mu)$ leptons (the gold-plated mode). Using the signal and background 
event rate, we obtain bounds on radion mass $m_\phi$ and radion vev $\vphi$ at
the $5\sigma$, $10\sigma$ discovery level.}

\vskip 5pT
\bfl
%{\it Keywords}: Brane world phenomenology; Z boson.\\
%\vspace*{0.05in}
\efl

\vskip 5pT

\newpage

\section{Introduction}
The Randall-Sundrum (RS) model~\cite{RS},  proposed as a resolution of the well-known 
electroweak hierarchy problem, is particularly interesting from the 
phenomenological point of view~\cite{GMPK}. According to this model, the
world  is $5$-dimensional and the extra spatial 
dimension is $S^1/Z_2$ orbifold. The metric describing such a world 
can be written as
\bea
d s^2 = e^{-2 k R_c |\theta|}\eta_{\mu \nu} d x^\mu d x^\nu
- R_c^2 d \theta^2,
\eea

\noindent
where $k$ is the bulk curvature constant and $R_c$ determines the size
of the extra dimension. The variable $\theta$ parametrizes extra dimension.
The model is constructed out of  
two $D_3$ branes which are located at two orbifold fixed points 
$\theta = 0$ and $\theta = \pi$, respectively. The brane located at  
$\theta = 0$ (where gravity peaks) is known as the Planck brane, while
the brane located at $\theta = \pi$ [where 
the Standard Model (SM) fields reside and gravity is weak] is known as 
the TeV brane. 
The factor $e^{- 2 k R_c |\theta|}$ appearing in the metric is known 
as the warp factor.
The length $R_c$, the distance between the two branes, can be 
related to the vacuum expectation value (vev) of some modulus
field $T(x)$ which corresponds to the fluctuations of the metric over the
background geometry given by $R_c$. Replacing $R_c$ by $T(x)$, we can rewrite
the RS metric at the orbifold point $\theta = \pi$ as
\bea \label{eqn:rsmetricVis}
d s^2 = g_{\mu \nu}^{vis} d x^\mu d x^\nu 
\eea

\noindent
where $g_{\mu \nu}^{vis} = e^{- 2 \pi k T(x)}\eta_{\mu \nu}
= \left(\frac{\phi(x)}{f}\right)^2 \eta_{\mu \nu}$. Here
$f^2 = \frac{24 M_5^3}{K}$ and $M_5$ is the $5$-dimensional Planck scale.
One is thus left with a scalar field $\hat{\phi}(x)$ 
[$\hat{\phi}(x) = \phi(x) - \vphi $] which is known as the radion 
field \cite{GRW}. 
Golberger and Wise  \cite{GW} proposed a mechanism for generating 
the potential for the field $T(x)$, with its minima at $R_c$
which satisfies $k R_{c} \simeq 11 \sim 12$, a requirement for the 
hierarchy resolution.
With this nonminimal version of the RS model with the radion $\hat{\phi}(x)$ 
stabilzed at $\vphi$ via the Golberger and Wise mechanism,
one can do a lot of interesting 
phenomenological studies. In particular, the radion, 
which can be lighter than the 
other low-lying gravitonic degrees of freedom in the RS model, 
will reveal itself first directly in the collider experiment
or indirectly in the precision measurement and verify our notion
of extra dimensions. Studies based on observable consequences 
of radion are available in the literature \cite{Cheung},\cite{DM}.

Here we want to look at the pair production of $Z$ bosons mediated by the
radion at large hadron collider (LHC). The 
crucial roles in enhancing the signal in this channel are 
($a$) the accessibility of the
radion resonance for $m_{\phi} > 2 m_{Z}$ and ($b$) the relatively
enhanced radion coupling (due to trace anomaly and top quark loop) with a 
pair of gluons at LHC energies, which we will discuss in succession in later 
sections. Finding the radion via the above channel, the so-called 
{\it gold-plated} mode, is already available in the literatures. However 
some error has been translated in those
studies due to the improper treatment of the Feynman rule in the 
radion coupling with a pair of off-shell top quarks. In the earlier studies, while estimating the radion production rate due to gluon fusion (i.e. in LHC context), radion coupling to a pair of top quarks is assumed to be proportinal to the mass $m_t$ of the top quark, which is definitely true if the produced top quarks are on-shell and obviously does not hold in the present case, since the top quarks are appearing inside the loop i.e. they are off-shell. This is the main difference between us and the already existed literatures and will be discussed in detail in 
Section 2.

The organization of the paper is as follows. In Sec. \ref{sec:section2}, we discuss the 
various interactions of the radion with the SM fields and determine the 
effective interaction of radion ($\phi$) with a pair of gluons ($g$) which is  due to the top loop and QCD trace anomaly. We discussed the difference 
in the production rate obtained by us with those alreay existed in the 
literatures. 
The general features of $Z$ boson  
pair production via radion are discussed in Sec.\ref{sec:section3}. 
Sec. \ref{sec:section4} 
contain discussions of the predicted signals for $m_\phi > 2 m_Z$, 
which is followed by the background estimate. We conclude in Sec. \ref{sec:section5}

%\newpage

\section{Effective interaction of radion with the SM fields}
\label{sec:section2}
Radion interactions with the SM fields confined 
on the TeV brane are governed by $4$-dimensional
general covariance. It couples to the trace of the
energy-momentum tensor of the SM fields in the following manner:
\bea \label{eqn:radcoup}
{\mathcal{L}}_{\it int} = \frac{\hat{\phi}}{\vphi} T^\mu_\mu (SM),
\eea
where $\vphi$ is the radion vev. The trace of the energy-momentum tensor of the SM fields is given by
\bea
T^\mu_\mu (SM) = \sum_{\psi} \left[\frac{3 i}{2} \left({\overline{\psi}}
\g_\mu \dnul \psi - \dnul{\overline{\psi}} \g_\mu \psi \right)\eta^{\mu\nu}
- 4 m_\psi {\overline{\psi}} \psi\right] - 2 m_W^2 W_\mu^+ W^{-\mu}
- m_Z^2 Z_\mu Z^\mu \nonumber \\
+ (2 m_h^2 h^2 - \partial_\mu h \partial^\mu h) + \cdots
\eea

\noindent The photon and the gluons couple to the radion 
via the usual top loop diagrams. {\it It is important to note at this 
point that as long as the fermions (produced from the decay of radion) 
are on-shell, their couplings 
are exactly same as the SM Higgs boson i.e. proportinal to the fermion mass and this makes it very difficult
to isolate radion from the Higgs boson by studying the final state 
decay products. 
However, if they are off-shell (e.g. radion production via top loop in gluons 
fusion), the situation is different due to the kinetic term of 
$\phi- t-{\bar{t}}$ coupling term in the kinetic energy part of the 
Lagrangian (see the discussion in section 2.1). Considering the correct 
Feynman rule for 
$\phi - t -\bar{t}$ coupling [ i.e. the kinetic term (missing in the literature) along with the mass term (present in the literature) ] and thus finding the correct 
radion production due to top loop in gluon fusion in the present context, differs crucially from those already existed in the literature} \cite{Cheung}.
Besides the top loop,
an added source of enhancement of radion production in gluon fushion is the QCD
trace anomaly, which we discuss below. For the sake of completeness, 
we first derive the lagrangian comprising radion-top coupling and then discuss the
effective radion-gluon coupling which is due to top loop and trace anomaly.

\subsection{Lagrangian for the radion-top coupling }
The radion coupling to the top quark in the Randall-Sundrum model
can be derived from the following action
\bea \label{eqn:Radtop}
S_1 = \int d^4 x \sqrt{- g_{vis}} \left[{\overline{\psi}}\left(i
\gamma_a e^{a\mu}D_\mu - m \right)\psi \right]. 
\eea
Note that the action (Eq.~(\ref{eqn:Radtop})) should also 
contain the Yukawa term 
$\frac{g_t}{\sqrt{2}} H {\overline{\psi}} \psi$ with $H$ being the SM 
Higgs field. Such a Yukawa term is irrelevant for the
present study and we will not consider this any further. The field 
$e^{a\mu}$ is the contravariant vierbein field for the visible brane and $g_{vis}$ is the determinant of the metric $g^{vis}_{\mu\nu}$ 
(Eq.~(\ref{eqn:rsmetricVis})). 
In the presence of radion fluctuation it satifies the normalization condition 
\bea
e^{a\mu}e_{a}^{\nu} = g^{\mu\nu} =
\left(\frac{\phi}{f}\right)^{-2}~\eta^{\mu\nu} =
e^{2 \pi k T(x)}~\eta^{\mu\nu}.
\eea
$D_\mu$ is the covariant derivative with respect to general coordinate
transformation and is given by 
$$D_\mu \psi = \partial_\mu \psi + \frac{1}{2}w_{\mu}^{ab} \Sigma_{ab}
\psi.$$ 
$w_{\mu}^{ab}$ is the spin connection and it can be computed 
from the vierbein fields by using the relation
\bea
w_{\mu}^{ab} = \frac{1}{2} e^{\nu a}(\partial_\mu e^b_\nu - \partial_\nu
e^b_\mu) - \frac{1}{2} e^{\nu b}(\partial_\mu e^a_\nu -
\partial_\nu e^a_\mu) -  \frac{1}{2} e^{\rho a}e^{\sigma b}(\partial_\rho
e_{\sigma c} - \partial_\sigma e_{\rho c})e^c_\mu, 
\eea
where $\Sigma_{ab}$ is given by the expression $\Sigma_{ab} = \frac{1}{4}
\left[\gamma_a,\gamma_b \right]$. In the presence of
radion fluctuations on the visible brane, the spin connection is given by
\bea
w_{\mu}^{ab} = \frac{1}{\phi} \partial_\nu \phi \left[e^{\nu
b}e^a_\mu - e^{\nu a} e^b_\mu\right].
\eea
The covariant derivative of the fermion field then becomes 
$$
D_\mu \psi = \partial_\mu \psi + \frac{1}{4
\phi}\partial^\nu \phi \left[\gamma_\mu,\gamma_\nu\right] \psi,
$$
where the $\gamma_\mu$ are independent of space time coordinates. The
action that determines the radion couplings to top quark can therefore be
written as
\bea \label{eqn:Actradtop}
S_1 = \int d^4 x \left(\frac{\phi}{f}\right)^4
\left[\left(\frac{\phi}{f}\right)^{-1} {\overline{\psi}}\{i
\gamma^\mu \partial_\mu + \frac{3 i}{2 \phi} \partial_\mu \phi
\gamma^\mu \}\psi - m_t {\overline{\psi}} \psi \right] \noindent
\nonumber \\
= \int d^4 x \left[{\overline{\tilde{\psi}}}\{i \gamma^\mu \partial_\mu
{\tilde \psi} + \frac{3 i}{2 \phi}\partial_\mu \phi \gamma^\mu
{\tilde \psi} \}\left(1 + \frac{\hat{\phi}}{\vphi}\right)^3 -
\tilde{m_t} \left(1 +
\frac{\hat{\phi}}{\vphi}\right)^4 {\overline{\tilde{\psi}}} {\tilde \psi}
\right] \noindent
\nonumber \\
= \int d^4 x \left[{\overline{\tilde{\psi}}}i \gamma^\mu \partial_\mu
{\tilde \psi} - \tilde{m_t} {\overline{\tilde{\psi}}} {\tilde{\psi}}
\right] 
%\noindent
%\nonumber \\
+  \int d^4 x \left[ \frac{3 i}{ \vphi} {\overline{\tilde{\psi}}}
\gamma^\mu \partial_\mu {\tilde \psi}~ {\hat{\phi}} + \frac{3 i}{2 \vphi}
{\overline{\tilde{\psi}}} \gamma^\mu {\tilde \psi}~ \partial_\mu
{\hat{\phi}} - 4 \tilde{m_t} 
\frac{\hat{\phi}}{\vphi}{\overline{\tilde{\psi}}} {\tilde
\psi}\right] \noindent
+ \cdots
\eea
%+  \int d^4 x \left[ 3~ {\overline{\tilde{\psi}}}i
%\gamma^\mu \partial_\mu {\tilde \psi}~ \frac{\hat{\phi}^2}{\vphi^2} + 
%\frac{3 i}{\vphi^2}~ {\hat{\phi}}~ {\overline{\tilde{\psi}}}\gamma^\mu
%{\tilde \psi}~ \partial_\mu {\hat{\phi}} - 6 \left(\tilde{m_t} 
%+ \frac{g_t}{\sqrt{2}} \tilde{H}
%\right)\frac{\hat{\phi}^2}{\vphi^2}{\overline{\tilde{\psi}}}
%{\tilde \psi}\right]
%\eea

We keep terms containing
one $\hat{\phi}$ field $[$ the second term in the last line 
of Eq.~({\ref{eqn:Actradtop}})$]$ which gives rise to the 
radion-top interaction with the following Feynman rule:
% ------------------------------------------------------------------
\begin{center} \label{fig:radtop}
\hspace*{0.25in}
\includegraphics[height=1.75in]{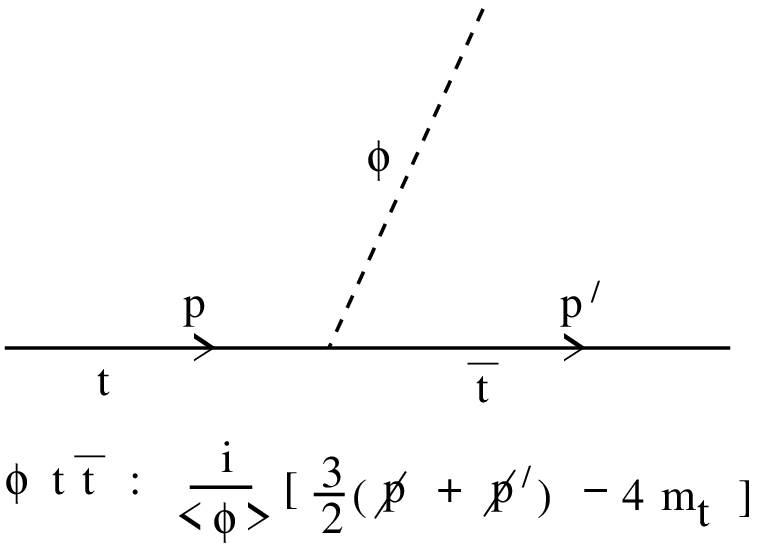}
\end{center}
\noindent {\bf FIG. 1}.
{\footnotesize\it Feynman diagrams for the radion coupling with a pair of 
top quarks.}
% ------------------------------------------------------------------
\bea 
\frac{i}{\vphi}\left[\frac{3}{2}(\slash{p} + \slash{p'}) - 4 m_t\right],
\eea 
where $p,~p'$ correspond to four-momenta of $t$ and $\overline{t}$ quarks  (see Fig. 1) and $m_t$, the top quark mass. 
Note that if the top quarks are on-shell, the above Feynman rule can be 
shown (using the equation of motion) to be equal to 
$-i \frac{m_t}{\vphi}$ similar to the corresponding 
SM Higgs boson coupling:
$-i \frac{m_t}{v}$ where the electro weak vev $v=247$ GeV. 
These two couplings are exactly identical 
in the limit $\vphi \to v$. 
In Eq.~(\ref{eqn:Actradtop}) the Ellipsis ($\cdots$) corresponds to terms 
involving two,three $\hat{\phi}$ fields [see \cite{Das} for the details]. 
Note that in the above $\psi = \left(\frac{f}{\vphi}\right)^{3/2} {\tilde{\psi}}$, and~ 
$m_t = \left(\frac{f}{\vphi}\right) {\tilde{m}}_t$. In the following, 
we shall assume that all fields and parameters have been
properly scaled so as to corresponds to the TeV scale and drop the
$\it{tilde}$ sign.

\subsection{Effective radion-gluon coupling:~QCD anomaly and top loop }
\subsubsection{QCD anomaly at order $\alpha_{s}$}
The radion couples to the trace of the energy-momentum tensor ($T_{\mu\nu}$), 
which can be equated with the four-divergence of the dilatation 
current ($D_\mu$) associated with scale symmetry of the theory i.e.
$$
\partial^\mu D_\mu = \partial^\mu (T_{\mu\nu} \delta x^\nu) = T^\mu_\mu.
$$
Scale symmetry is a good symmetry at the classical level (i.e. $\partial^\mu D_\mu = T^\mu_\mu = 0$) in the limit of zero mass and in the absence of 
dimensionful couplings. However, quantum correction breaks the scale 
invariance and the divergence of the dilatation current receives the anomalous contribution given by $\partial^\mu D_\mu = T^\mu_\mu \neq 0$. 
Such a non-zero anomalous $T^\mu_\mu$ generated due to the quantum correction 
(zero at the classical level), is known as the 
{\it Trace Anomaly} or {\it Weyl Anomaly} term and it can be shown to be 
proportional to the relevant beta function of the underlying gauge theory
\cite{CDJ}.
The radion coupling to the massless gluons (or photons) 
is crucially controlled by such a trace anomaly term. For gluons, 
the interaction lagrangian due to this trace anomaly term can be written as 
\bea
T^\mu_\mu (SM)^{anom} = \sum_{a} \frac{\b_s(g_s)}{2 g_s} G_{\mu\nu}^a 
G^{a \mu\nu},
\eea
where $g_s$ is the strong coupling constant and 
$\b_s (g_s)/{2 g_s} = - [\a_s/{8 \pi}]~ b_{QCD}$. Here 
$b_{QCD} = 11 - 2 n_f/3$ and $n_f$ is the number of quark flavours.

\subsubsection{Top loop at order $\alpha_{s}$}
At order $\alpha_s$, a radion can couple to a pair of gluons via the one-loop
diagrams of Fig. 2. We follow the usual \cite{ABJ} procedure to evaluate
the amplitude for the process $\phi \goes g g $. Consistency 
with the Lorentz 
invariance and the transverse nature of the gluon 
[i.e. $P^\mu \varepsilon^{a*}_\mu(P) = Q^\nu \varepsilon^{a*}_\nu(Q) = 0$] 
leads to the following form  of the amplitude:
\bea \label{eqn:ggr1}
{\cal M}\left[\phi(k) \to g(P) g(Q)\right] &=& M^{\mu \nu} 
\varepsilon^{a*}_\mu(P)\varepsilon^{b*}_\nu(Q) \noindent 
\nonumber \\
&=& \left[ F_1(k^2) P^\nu Q^\mu + F_2(k^2) \eta^{\mu \nu} \right]
\varepsilon^{a*}_\mu(P)\varepsilon^{b*}_\nu(Q)
\eea
with $\eta_{\mu\nu} = diag(1,-1,-1,-1)$ and $k=P+Q$.
Imposition of the current conservation $Q_\nu M^{\mu \nu} = 0$ (follows from
the gauge symmetry) relates the two form factors as
\bea
F_2 (k^2) = -  P.Q  ~F_1 (k^2).
\eea
Inserting this in Eq.(\ref{eqn:ggr1}), we find the 
amlitude ${\cal M}(\phi \to g g)$ as
\bea \label{eqn:ggr2}
{\cal M}(\phi \to g g) =
F_1(k^2) \left[ P^\nu Q^\mu - P.Q \eta^{\mu\nu} \right]
\varepsilon^{a*}_\mu(P)\varepsilon^{b*}_\nu(Q).
\eea
We now evaluate the set of diagrams of Fig. 2 and find $F_1(k^2)$ as
\bea
 \frac{\a_s(\mu^2) \delta_{ab} Q^2_f}{2 \pi \vphi} I_{QCD}, 
\eea
where $Q_f$ is $2/3$ for the top quark. Diagrams 2(c) and 2(d) do not give rise 
to contributions of the form Eq.~(\ref{eqn:ggr2}), a necessary requirement 
 which follows from the gauge invariance and other constraints (discussed above). 
Thus their contributions to the effective $\phi-g-g$ interaction, are simply 
zero. 
Only diagrams 2(a) and 2(b) contribute.
%\newpage
%%\vspace*{-1.5in}
% ------------------------------------------------------------------
\begin{center} \label{fig:toploop}
\includegraphics[height=3in]{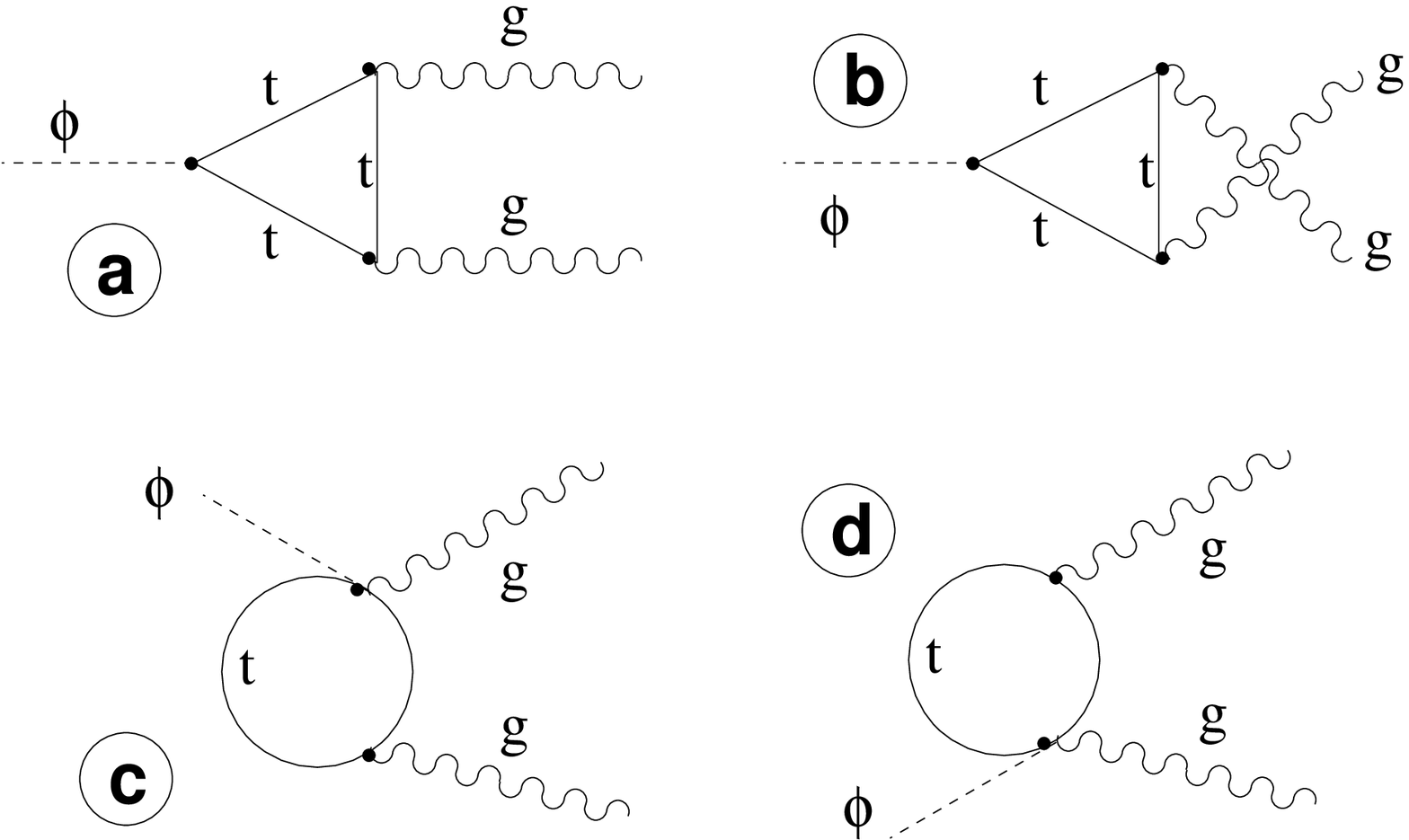}
\end{center}
\noindent {\bf FIG. 2}.
{\footnotesize\it Feynman diagrams with a top quark loop contributing to the 
process $\phi \rightarrow g g $ at order $\alpha_s$.}
% ------------------------------------------------------------------

 The loop function $I_{QCD}$ arising in the process is given by
\bea \label{eqn:IQCD}
I_{QCD} &=&  -\left[ 20 - 12 \sqrt{x_t - 1} ~\l(x_t) - 2 (4 x_t - 1)\l^2 (x_t) 
\right]~~for ~x_t > 1, \nonumber \\
&=&  -\left[ 20 - 12 \sqrt{1 - x_t }~ \l(x_t) - 2 (1 - 4 x_t)\l^2 (x_t) \right]~~for ~x_t \leq 1,
\eea
where $x_t = 4 m_t^2/m^2_\phi$ and $m_\phi$ is the radion mass. Note the 
presence of the constant term and the term linear in $\l(x_t)$ arising in the 
loop function. They, found to be absent in the literature 
\cite{Cheung}, arises due to the presence of the kinetic term 
in the $\phi -t -\bar{t}$ coupling. As we will see, this results in an 
important role in the radion production rate.
The function $\l(x_t)$ is given by
\bea \label{eqn:lambda}
\l(x_t) &=& tan^{-1} \left(\frac{1}{\sqrt{x_t - 1}}\right)~~for ~x_t > 1,\nonumber \\
&=& \frac{1}{2}\left[ln\left(\frac{1 + \sqrt{1 - x_t }}{ 1 - \sqrt{1 - x_t }}\right) - i \pi \right] ~~ for ~x_t \leq 1.
\eea
Note that for $x_t > 1$, $n_f = 5$, so 
$b_{QCD}$ [arising in the trace anomaly term (see Sec. 2)] is $23/3$, while 
for $x_t \leq 1$, $n_f = 6$ and $b_{QCD} = 21/3$. Note that the $x_t \le 1$ i.e. $m_\phi \ge 2 m_t$, corresponds to the top resonance inside the loop and is the region where the heavy quark appriximation breaks down.

 {\it On the whole}, the effective 
$\phi-g-g$ interaction finds contributions from the trace anomaly and the 
top quark loop and hence, can be written as  
\bea
\frac{i \a_s(\mu^2) \d_{ab}}{2 \pi \vphi} \left[ b_{QCD} + Q^2_fI_{QCD}  
\right]
\left[ P^\nu Q^\mu - P.Q \eta^{\mu\nu} \right]. 
\eea
With the above interactions, we calculate the decay width $\Gamma [\phi \to gg]$ as
\bea
\Gamma [\phi \to gg] = \frac{\a_s^2(\mu^2) m_\phi^3}{32 \pi^3 \vphi^2} |b_{QCD} + Q^2_f I_{QCD}|^2.
\eea
Radion decay to other SM fields have been discussed in Refs.~\cite{Cheung, DRR}. 

\section{Neutral $Z$ boson pair production: LHC prospects }
\label{sec:section3}
The neutral $Z$ boson pair production from gluon-gluon fusion 
at the LHC has been  extensively
studied in the context of the SM  as well as 
several of its extensions such  as the minimal supersymmetric standard 
model (MSSM).  
A precise estimate for the $Z$  
boson pair production due to Higgs resonance and a detailed 
analysis of it's subsequent decay into
charged leptonic final states is available in the literature \cite{AD}. 
Any excess over this can be interpreted as the signature of new physics 
\cite{GBM}.
The mediation of the heavier neutral Higgs in supersymmetry has been shown
to be the source of some enhancement in a region of the parameter space.
%\cite{GBM} 
Theories with extra dimensions, too, have been explored in this context, both
in the ADD (Arkani-Hamed, Dimopoulos and Dvali \cite{ADD}) and RS models \cite{PSS}. 
It has been reported in these papers
that the mediation of gravitons can boost the $Z$ boson pair production rates.
In this paper we wish to emphasize that the presence of a stablized 
radion in the nonminimal RS context is particularly significant. This is because 
(i) {\it whereas the graviton resonance in the RS scenario is usually at too
high a mass to be significant, a relatively less massive radion can be
within the kinematic reach}, and also 
(ii) {\it the enhancement of radion coupling to
a pair of gluons via the trace anomaly term and top loop 
picks up the contributions}.
We have computed the predicted rates for the case $m_\phi > 2m_Z$
and analyzed the viability of the
resulting signals with appropriate event selection strategies.

\noindent Using Breit-Wigner 
approximation for the resonant production of a radion, one finds the 
following expression for the cross section for $p p \rightarrow \phi 
\rightarrow Z Z$  (the dominant partonic subprocess being 
$g g \rightarrow \phi \rightarrow Z Z$):
\bea \label{eqn:radprod1}
\sigma_{ZZ}(p p \rightarrow \phi \rightarrow Z Z) = \int d x_1 \int d x_2~ 
g_1(x_1,Q^2)~ g_2(x_2,Q^2) \hat{\sigma}_{ZZ}(g g \rightarrow \phi \rightarrow Z Z),
\eea
where
\bea \label{eqn:radprod2}
\hat{\sigma}_{ZZ}(g g \rightarrow \phi \rightarrow Z Z) &=&
~\frac{\a^2_s(\mu^2) }{2048~\pi^3 \vphi^4} 
|b_{QCD} + Q^2_f I_{QCD}|^2~ \frac{\hat{s} [\hat{s}^2 - 4 m_Z^2\hat{s} + 12 m_Z^4 ]}
{[(\hat{s} - m_\phi^2)^2 + m_\phi^2 \Gamma_\phi^2]} \nonumber \\
&=&
~\frac{\Gamma[\phi \rightarrow gg]}{64~m_\phi^3 \vphi^2}
~\frac{\hat{s} [\hat{s}^2 - 4 m_Z^2\hat{s} + 12 m_Z^4 ]}
{[(\hat{s} - m_\phi^2)^2 + m_\phi^2 \Gamma_\phi^2]},
\eea
In obtaining Eq.(\ref{eqn:radprod2}), we made use the decay width of 
radion ($\phi$) into a pair of gluons ($g$) i.e. $\Gamma [\phi \to gg]$ and 
assumes the validity of time reversal symmetry.
Here $\hat{s}$ ($= x_1 x_2 s$) is the centre-of-mass energy for the partonic 
subprocess, ${\sqrt{s}} (=14$~TeV) corresponds to the proton-proton
centre-of-mass energy, $g_1(x_1,Q^2)$ and 
$ g_2(x_2,Q^2) $ are the gluon distribution functions of the two 
colliding protons,~$x_1$ and $x_2$ are the momentum 
fractions of the partons (gluons) of the colliding protons and 
$\mph$ and $\Gamma_\phi$ denotes the mass and total decay width of the 
radion.  The factorization scale $Q^2$ and renormalization scale
$\mu^2$ appearing in $g(x,Q^2)$ and $\a_s (\mu^2)$  
are fixed at $Q^2 = \mu^2 = \hat{s}$ in our analysis.
We have used  CTEQ4L \cite{CTEQ} parton distribution functions, setting the 
renormalisation scale at the partonic subprocess centre-of-mass energy 
($\sqrt{\hat{s}}$). We have checked that the predicted
results are more or less unchanged if this scale is set at radion mass 
($m_\phi$).

 In Eq.~(\ref{eqn:radprod2}), we find that away from the radion resonance, 
the term proportional to  $\Gamma^2_\phi$
is of little consequence, and the total rate falls as  $1/\vphi^4$.
This is because the decay width of the
radion in any channel is proportional to $1/\vphi^2$. 
Near resonance, on the other hand,
the contribution is dictated by the term with $\Gamma_\phi$ in the
radion propagator. Since, in the narrow width approximation, one can write
$1/[(\hat{s} - m_\phi^2)^2 + m_\phi^2 
\Gamma_\phi^2] \simeq \pi \delta(\hat{s} - m^2_{\phi})/
m_{\phi} \Gamma_{\phi}$, the overall rate falls as  $1/\vphi^2$ near 
resonance.

%\newpage 
\section{Numerical Analysis}
\label{sec:section4}
\subsection{Event selection and results: $m_\phi > 2 m_Z$}

The signal of $Z$-pair production due to radion resonance depends on the final 
states produced by the decay of $Z$-boson. In our analysis, we consider the 
mass range $m_\phi > 2~m_Z$ and in such a case, the $BR(\phi \to b \bar{b})$
 falls to $0.0028$ i.e. $\sim 3\%$ (say at $m_\phi = 200$ GeV) and hence the 
 $b \bar{b}$ mode for detecting the radion resonance, is not worthwhile to 
look. 
On the other hand for the same $m_\phi$ value, the $BR(\phi \to ZZ)$ becomes 
$13\%$, and can be considered as quite useful for detecting 
the radion resonance. We will consider the scenerio in which the 
produced $Z$ will decay 
mainly leptonicaly (i.e. $l = e, \mu $). So, the final state contains 
$4$ charged leptons and normally such events may have appreciable SM 
background arising from the following sources:

\begin{itemize}
\item The electro weak production of $l^+ l^- l^+ l^- (l= e, \mu)$,
\item Higgs production due to gluons fusion, which after subsequent decay into
$2~Z$ bosons can produce $4$ charged leptons in the final state. 
This is the dominant background process. 
\end{itemize}
However, using a proper set of cuts one can diminish this 
background effect, thereby, enhancing the signal. The cuts used in our 
analysis can be listed as follows:
\begin{itemize}
\item Each of the produced charged lepton ($e$ or $\mu$ type) must have the
 pseudorapidity $\eta$ satisfying $-3 < \eta < 3$.
\item  Each of the produced charged lepton must have transverse momenta $p_T$ 
greater than $15$ GeV.
\end{itemize}

\noindent After applying all the above cuts, the cross section for the process 
$pp(gg) \rightarrow \phi \rightarrow Z Z \rightarrow l^+ l^- l^+ l^-$ 
(the signal)
and $pp(gg) \rightarrow h \rightarrow Z Z \rightarrow l^+ l^- l^+ l^- $
and $pp(gg) \rightarrow (EW, non-resonant) \rightarrow l^+ l^- l^+ l^- $ 
(the backgrounds) are obtained. 
In Fig. 3 we have plotted these cross sections for the signal and background
against $m_S$ (GeV). For the signal $m_S = m_\phi$ and 
$\vphi = 250,~500,~750,~1000$ GeV with the SM Higgs mass  
$m_h = 185$ GeV. For the background $m_S = m_h$ for the resonating Higgs channel, besides other non-resonating electro weak channels.
 We considered the cases in which $Z$ decay into electrons 
and muons ($l=e,\mu$) with total 
branching ratios $0.0682$. In 
addition, an average detection efficiency of $80 \%$ per lepton has been 
assumed. Below, we make the following observations 
\bi
\item From Fig. 3, we see that there is a substantial 
enhancement of the rates (signal) in comparison to what is obtained in the 
Standard Model. The bump at around $M_S = 350$ GeV, resembles the top
resonance inside the top loop (the scalar (radion/Higgs) 
production channel via gluon fusion). Such a top resonance, encoded in the 
function $\l(x_t)$ (see Eq.(\ref{eqn:IQCD})) and appearing in the loop 
integral, 
seems to happen at $m_S \ge 2 m_t$  which is being manifested 
in the imaginary part of the $\l(x_t)$ function (Eq.(\ref{eqn:lambda})).

An estimate of the signal and background event rate (obtained by 
multiplying the cross section with the integrated luminosity) is given in Table 1 and 2.

%\newpage
%\vspace*{-0.5in}
%-----------------------------------------------------------------------
\begin{figure}[htb] \label{fig:sigbgd}
\begin{center}
\vspace*{4.5in}
      \relax\noindent\hskip -7.4in\relax{\includegraphics{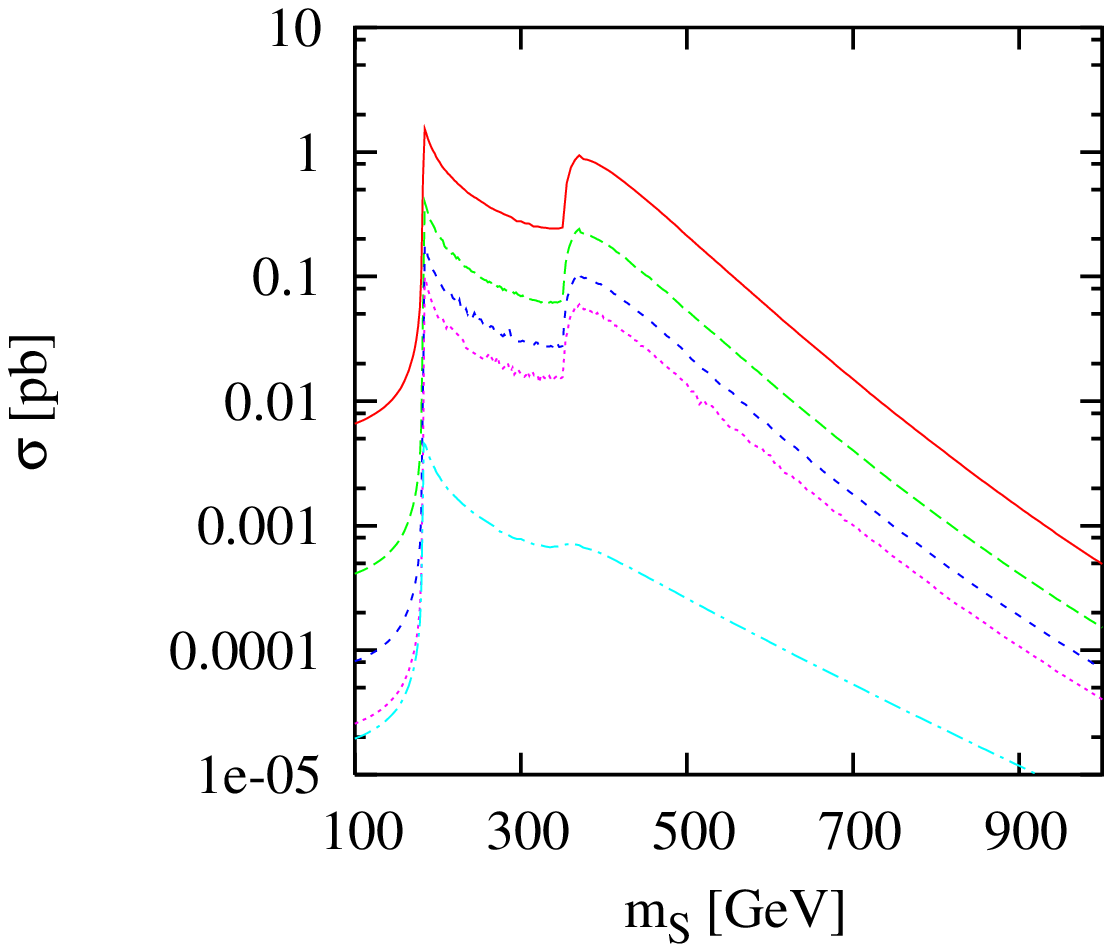}}
\end{center}
\end{figure}
\vspace*{-0.95in}
\noindent {\bf FIG. 3} (color online):
{{ \it Plots showing the total cross section of 
$4~l (l = e, \mu)$ in pp collisions due to radion resonance as a function of 
the scalar($S$) mass $m_S$, with $S$ is $\phi$~and $h$ for the signal and the background, respectively. For the signal, we choose the SM Higgs 
mass $m_h =185$ (GeV) and $\vphi$ = 250~(solid curve),~500~(long-dashed   
curve),~750~(short-dashed curve) and 1000~(dotted curve) (GeV). The lower (dashed-dotted) curve  
corresponds to the SM background. We choose  
$Q^2 = \mu^2 = \hat{s} $, although the result remains more or less unchanged if one chooses $Q^2 = \mu^2 = m_\phi^2 $.}}
%--------------------------------------------------------------------- 

\item From  Fig. 3 and Table 1, it is clear that the total 
cross section (and hence the event rate) of the signal falls with the increase 
in $\vphi$, i.e. near resonance it goes as 
$\frac{1}{\vphi^2}$ while away from the resonance it goes as 
$\frac{1}{\vphi^4}$ (see also the discussion of Sec. 3).
 
\item Figure 3 (see also Table 1)  shows that, for a given  $\vphi$, 
the cross section falls with the increase 
in $m_\phi$ [see Eqs.(\ref{eqn:radprod1}) and (\ref{eqn:radprod2})].

\item  The total event(signal) rate is greater than that 
predicted in ADD and RS scenerio involving spin-2 gravitons. As has been 
mentioned before that this enhancement is due to two reasons, namely, (a) 
possibility of having a radion resonance and (b) the enhanced coupling at 
radion-gluon-gluon vertex due to trace anomaly and top loop. It is worthwhile
to give some number related to the signal cross section which differs from the literatures, say, Chaichian {\it et. al.} in the $\xi=0$, where $\xi$ is 
the radion-Higgs mixing parameter \cite{Cheung} and as we mentioned earlier that such difference is due to the treament of the $\phi-t-\bar{t}$ coupling required in the traingle top loop evaluation. 
For example, below $2 m_t$ threshold, say at around $m_\phi = 200$ GeV, we find $\sigma = \sigma_{s}= 0.047$ pb in contrast to $\sim 0.01$ pb (Chaichian {\it et. al}) for the radion vev $\vphi$  about $1$ TeV. Above $2 m_t$ threshold, with the same $\vphi$ value, say at $m_\phi = 370$ GeV, we find the 
signal cross section 
$\sigma$ about $0.06$ pb, whereas they found $0.025$ pb. 
\ei 

\begin{table}[htb]
\caption{Showing the event (signal) rate as a function of radion mass $m_\phi$ and vev $\vphi$.}
\begin{center}
%\begin{ruledtabular}
\begin{tabular}{lllrrrr}
\hline
\hline
$m_\phi(GeV)$ & $\vphi$ (GeV) &  $\sigma_s (fb)$ & $\cal L$ ($fb^{-1}/yr$) & Events/yr\\
\hline
    & 250 & 408 & 100 &  40800\\ 
 250& 500 & 100 & 100 & 10000 \\ 
    & 750 & 45 & 100 & 4500 \\ 
    & 1000 & 22& 100 & 2200 \\ 
    & 250 & 213 & 100 & 21300\\ 
 500& 500 & 53 & 100 & 5300 \\ 
    & 750 & 24 & 100 & 2400 \\ 
    & 1000 & 14 & 100 & 1400 \\ 
\hline
\hline
\end{tabular}
\end{center}
%\end{ruledtabular}
\end{table}

\begin{table}[htb]
\caption{Background(bgd) event rate estimation with the assumption that the dominant contribution comes from the SM Higgs (with $m_h = 185$ GeV) resonance due to gluon-gluon fusion. }
\begin{center}
%\begin{ruledtabular}
\begin{tabular}{lllrrrr}
\hline
\hline
$\sigma_B (fb)$  & $\cal {L}$ ($fb^{-1}/yr$) & Events(bgd)/yr\\
\hline
 4.4 & 100 & 440 \\
\hline
\hline
\end{tabular}
\end{center}
%\end{ruledtabular}
\end{table}

%---------------------------------------------------------------------  
\vspace*{0.1in}
\bi
\item  A bump at around $m_\phi = 350$ GeV corresponds to the top resonance inside the top loop. Above this resonance (the $2 m_t$ threshold), the cross section is mainly dominated by the top loop.  Note that, before this resonance, the large signal cross section 
was mainly due to trace anomaly, which remains important even
after this $2 m_t$ threshold. 
\item While maintaining the fact that the signal cross section 
be greater than that of the SM background, one can even allow larger $m_\phi$ 
values corresponding to lighter $\vphi$ choices, the feature which will be
explored in the next section. 
%For heavier $\vphi$ values,  the upper cut on $m_\phi$ values gets lowered. 
\ei
\subsection {Significance contours}
 With an integrated luminosity of $100 ~fb^{-1}$, the above rate indicate a 
rather impressive prospect of detecting the radion resonance via the 
pair-produced $Z$ bosons and it's subsequent charged leptonic decays. 
To gauge the situation, however, one should 
also remember that the backgrounds, even after applying proper set of cuts, 
are never totally eliminated and one 
can compare the signal with this background. In Fig. 4, we show the
contour plots in the plane of radion mass $m_\phi$ and radion vev 
$\vphi$ corresponding to 
$\frac{S}{\sqrt{B}} (= R_{SB})=5,~10$, 
where $S$ and $\sqrt{B}$ are the number of events corresponding to the signal 
and the background with the above luminosity.
In calculating the backgrounds, we have taken both statistical and
systematic effects into account, assuming the systematic uncertainty to be
$2\%$ of the total background and adding it in quadrature to the computed
background itself. For each curve in Fig. 4, the region 
above the curve is allowed. For example, for an intermediate (heavy) 
$m_\phi$, say about
$250$ ($500$) GeV , we find lower bounds on $\vphi$ about $\sim 2741$ ($778$) GeV  corresponding to $R_{SB} = 5$ and $\sim 2076$ ($\sim 556$) GeV 
corresponding to 
$R_{SB} = 10$.  These plots corresponds to finding 
the signal over a large region of the parameter space and 
 suggests the possibility of having a heavier radion mass with a moderately 
lower radion vev $\vphi$, although the possibility of having an intermediate radion mass is not ruled out.

%\vspace*{-1.5in}
%-----------------------------------------------------------------------
\begin{figure}[htb]
\begin{center}
\vspace*{4.5in}
      \relax\noindent\hskip -7.4in\relax{\includegraphics{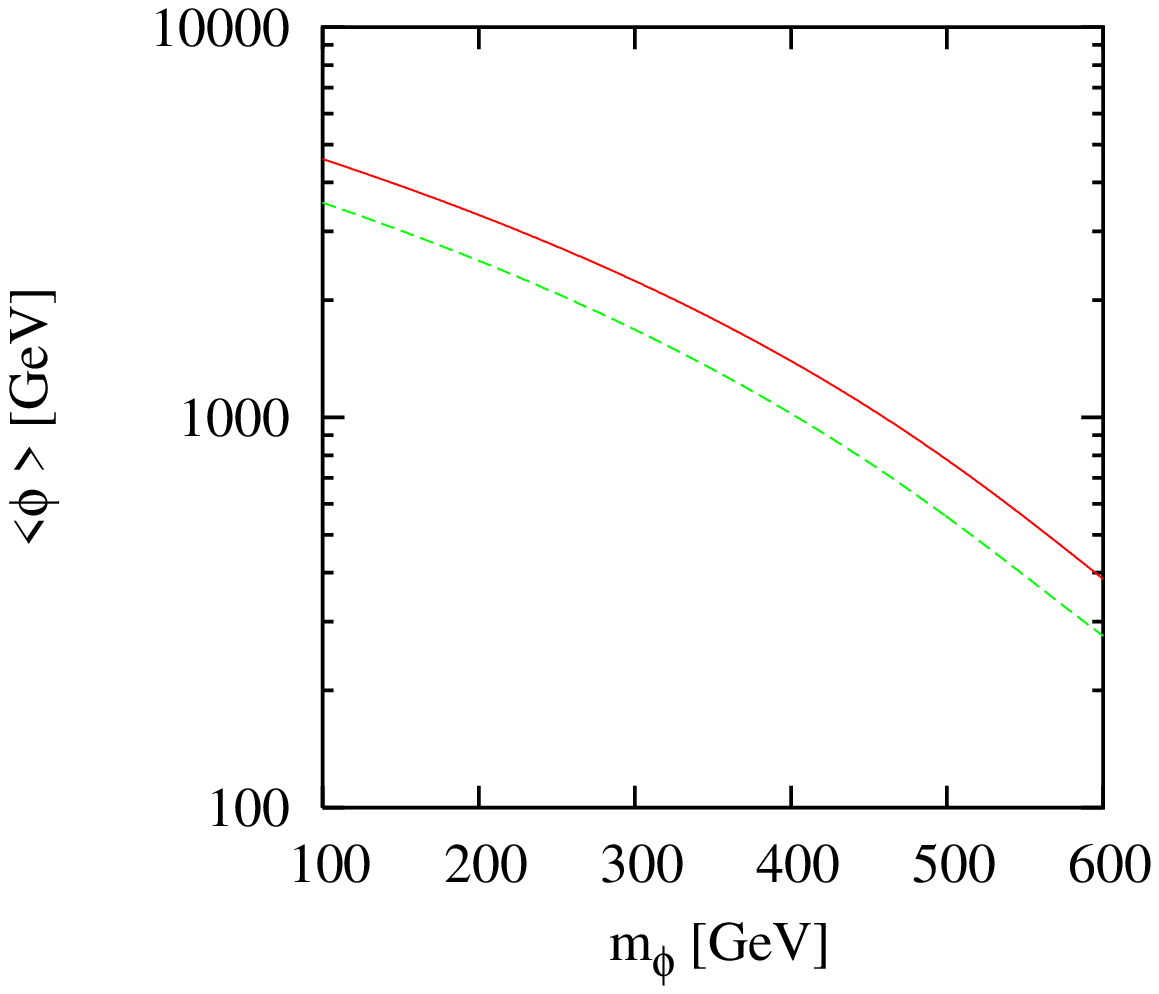}}
\end{center}
\end{figure}
\vspace*{-1.0in}
\noindent {\bf FIG. 4} (color online).
{\it Contour plot in the plane of $\mphi$ and $\vphi$ corresponding to
$R_{SB}$ = 5~(solid line) and 10~(dashed line). }
%---------------------------------------------------------------------

\section{Summary and Conclusion}
\label{sec:section5}
The enhanced coupling of the radion-gluon-gluon due to top loop and trace 
anomaly offers an immense possibility of finding
radion at Large Hadron Collider. We have considered the correct radion-top-antitop coupling and fix the radion production rate at LHC. We have considered the channel in which the radion 
 is produced as a resonance 
in gluon-gluon fusion and then decay to a pair of  
$Z$ bosons, which subsequently decays 
into $4$ charged leptons, $l(=e,\mu)$. The SM background comprising 
$4~l$ final sets is found to be 
smaller that that due to the signal (radion resonance) even after applying a proper
set of cuts for a wide range of radion mass ($m_\phi$) values. 
For $m_\phi$ ranging from $180$ to $1000$ GeV,
the signal cross section dominates over that due to SM background. 
Exploiting the signal and background ratio, we make the contour plots in 
$m_\phi - \vphi$ plane. The contour plot suggests the 
possibility of having an intermediate to heavier
radion of mass about $\sim 250~(500)$ GeV with $\vphi$ about 
$2076~(556)$  GeV at the $10\sigma$ discovery level. 

\bfl
{\Large {\bf Acknowledgement}}
\efl
I thank Prof. U.~Mahanta (deceased) and Prof. S.~Raychaudhuri 
for introducing me to brane world phenomenology and Prof. Pankaj Jain for 
his very useful comments and suggestions after reading this manuscript. 
Special thanks are reserved for Prof. Kwei-Chou Yang who provided  
me a nice stay with NSC support at CYCU, Taiwan, where this work 
was finally completed.

%\newpage

\end{document}